\documentstyle[12pt]{article}
\newcommand{\be}{\begin{equation}}
\newcommand{\ee}{\end{equation}}
\newcommand{\bea}{\begin{eqnarray}}
\newcommand{\eea}{\end{eqnarray}}
\begin{document}
\title{Critical coupling in (1+1)-dimensional light-front $\phi^{4}$ theory
  }
\author{Kazuto Oshima\thanks{E-mail: oshima@nat.gunma-ct.ac.jp} and  
        Masanobu Yahiro\thanks{E-mail: yahiro@nat.gunma-ct.ac.jp} \\ \\
\sl Gunma National College of Technology, Maebashi 371-8530, Japan }
\maketitle
\begin{abstract}
The spontaneous symmetry breaking in (1+1)-dimensional 
$\phi^{4}$ theory is studied with 
discretized light-front quantization, that is, by solving 
the zero-mode constraint equation. 
The symmetric ordering is assumed for the operator-valued 
constraint equation. 
The commutation relation between the zero mode and each 
oscillator mode is calculated with $\hbar$ expansion.  
A critical coupling evaluated from the first some terms in the
expansion is $28.8\mu ^{2}/\hbar \le \lambda_{cr} \le 31.1\mu ^{2}/\hbar$ 
consistent with the equal-time one
$22\mu ^{2}/\hbar \le \lambda_{cr} \le 55.5\mu ^{2}/\hbar$.
The same analysis is also made under another operator
ordering.
\end{abstract}
\baselineskip=24pt
\newpage
The light-front formulation of field theory is formally 
equivalent to the equal-time formulation 
within the framework of perturbation theory \cite{Chang}. 
Nonperturbative calculations in the light-front formulation are
successful in lower-dimensional 
field theories and yield the correct particle spectra 
\cite{Brodsky,Burkardt,ours}. The light-front formulation is then 
expected to be a promising method for solving 
the nonperturbative dynamics such as QCD. 
Nevertheless, the application to spontaneous symmetry breaking (SSB) 
is obstructed by the unsolved problem of the vacuum structure 
on the light cone \cite{MY,Yamawaki,Hornbostel}. 
It is then an important issue
to confirm that for SSB the light-front formulation yields  
the same result as the equal-time formulation.

We study spontaneous breaking of the $Z_2$ symmetry in (1+1)-dimensional
$\phi ^{4}$  theory as an instance. 
It is well known in the equal-time formulation that 
the model undergoes a phase transition at strong coupling \cite{PT}.
In light-front field theory, the true vacuum always equals to 
the Fock vacuum \cite{MY}, 
so SSB is believed to occur through the zero mode;
a non-zero vacuum expectation value of the zero mode signals SSB. 
To ascertain this idea, several authors 
\cite{Heinzl,Harin,Pinsky1,Pinsky2,Xu}
have computed a vacuum expectation value of the
zero mode and a critical coupling 
with either a truncation of Fock space or perturbation. 
Their results are qualitatively satisfactory. However, 
it is not clear whether the prescriptions are  
reliable for nonperturbative phenomena such as SSB, and 
it is not easy to improve the accuracy of the calculations. 
Our purpose in this brief report is to propose 
a different computation rule and obtain more accurate results. 
This rule yields a systematic way of obtaining 
a critical coupling $\lambda_{cr}$ and a critical exponent $\beta$.  

We define the light-front coordinates $x^{\pm}=(x^{0}\pm x^{1})/\sqrt{2}$.
The Lagrangian density of  $(1+1)$-dimensional $\phi ^{4}$ theory
is described as
\be
{\cal L}=\partial_{+}\phi\partial_{-}\phi-{\mu ^{2} \over 2}\phi ^{2}
-{\lambda \over 4!}\phi ^{4}. 
\ee
We put the quantum system in a box of length $d$ and impose periodic boundary
conditions. The field operator $\phi$ is then expanded as
\be
\phi (x)={1 \over \sqrt{4 \pi}}a_{0}+
\sum _{n \ne 0}{1 \over \sqrt {4\pi |n|}}a_{n}(x^{+})e^{ik_{n}^{+}x^{-}},
\ee
where $k_{n}^{+}=2\pi n /d$.  
Coefficients of the expansion are operators which  
satisfy the canonical commutation relations  \cite{MY,Pinsky1,Pinsky2},
\be
[a_{k},a_{l}]=[a_{k}^{\dagger},a_{l}^{\dagger}]=0, \hspace{5mm}
[a_{k},a_{l}^{\dagger}]=\hbar \delta _{k,l}, \hspace{5mm}
k,l > 0,
\label{commutation}
\ee
where $a_{k}^{\dagger}=a_{-k}$. 
The zero-mode operator $a_{0}$ is not an independent quantity, since 
it governs the constraint equation \cite{MY,Pinsky1}
\bea
0&=&a_{0}^{3}+ga_{0}+\sum _{n \geq 1}{1 \over n}(a_{0}a_{n}a_{n}^{\dagger}
+a_{0}a_{n}^{\dagger}a_{n}+a_{n}a_{n}^{\dagger}a_{0}+
a_{n}^{\dagger}a_{n}a_{0}+a_{n}a_{0}a_{n}^{\dagger}+
a_{n}^{\dagger}a_{0}a_{n}\nonumber \\
& & - 3\hbar a_{0}) +\sum _{k,l,m \ne 0}
{\delta_{k+l+m,0} \over \sqrt{ |klm|} }a_{k}a_{l}a_{m} 
\equiv \Phi (a,a^{\dagger}) ,
\label{constraint}
\eea
where we have added the term $\sum _{n \geq 1}{-3 \over n}\hbar a_{0}$
to remove the tadpole divergences. 
The commutation relations (\ref{commutation}) and 
the constraint equation (\ref{constraint}) are obtained  
by quantizing the classical system with 
the Dirac-Bergmann quantization procedure. The procedure
does not specify an operator ordering 
in the constraint equation (\ref{constraint}), 
so we have assumed a Weyl (symmetric) ordering 
\cite{Pinsky3}. This assumption is reasonable, since it describes 
the phase transition properly, as shown later.
The zero mode $a_{0}$ is a complicated operator 
obeying Eq.(\ref{constraint}). 
We see from the quantization procedure  
that the zero mode satisfies \cite{Pinsky1}
\be
[a_{0},a_{m}^{\dagger}a_{m}]=0.
\label{conservation}
\ee
The commutation relation guarantees that $a_{0}$ preserves the
longitudinal momentum conservation. 

So far the operator relation $[a_{0},a_{n}]$ has not been 
esteemed significant. In literature, the corresponding classical 
relation (Dirac bracket) is derived from a classical constraint
equation  corresponding to Eq.(\ref{constraint}), but 
it is impossible to quantize the Dirac bracket  
by simply replacing it by the commutation relation.
Because, the quantization procedure does not specify 
an operator ordering for operator products appearing 
in $[a_{0},a_{n}]$. 
However, the importance of $[a_{0},a_{n}]$ should be emphasized. 
In fact, as soon as $[a_{0},a_{n}]$ is obtained, 
we can calculate $\sigma \equiv \langle 0|a_{0}|0 \rangle $
as a function of $g \equiv {24\pi \mu ^{2} \over \lambda  }$, 
as shown later. 
We derive the $[a_{0},a_{n}]$ 
with Eq.(\ref{conservation}) from 
\be 
[\Phi (a,a^{\dagger}),a_{n}]=0 , 
\label{derivation}
\ee
expanding $[a_{0},a_{n}]$ as
\be
[a_{0},a_{n}] \equiv \sum_{p \ge 1}{\tilde \alpha}_{p} (n)\hbar ^{p}.
\label{expansion}
\ee
If $[a_{0},a_{n}]$ is obtained without the expansion, 
it should be a function 
of $a_{0}$, $a_{n}$, $a_{n}^{\dagger}$ and $\hbar$. 
Expanding the operator-valued function, 
we see that the expansion series 
is not unique because of the ambiguity on operator ordering.
We only obtain ${\tilde \alpha}_{p}$ uniquely, when 
we take an operator ordering for ${\tilde \alpha}_{p}$. 
Since there would be no a priori proper operator-ordering for  
${\tilde \alpha}_{p}$, we take a simple operator ordering
shown below. 
Difference in operator ordering for ${\tilde \alpha}_{p}$ 
affects only to the higher order terms  
${\tilde \alpha}_{q} \quad (q > p )$.

At $O(\hbar)$, we have 
$[a_{0}^{3},a_{n}]=3a_{0}^{2}{\tilde \alpha}_{1} (n) \hbar  +O(\hbar ^{2})$
etc., and from Eq.(\ref{derivation}) we obtain
\be
{\tilde \alpha}_{1} (n)={6 \over nA}a_{0}a_{n}+{3 \over A}\sum _{k \ne 0,n}{1 
\over \sqrt{|kn(n-k)|} }a_{k}a_{n-k} ,
\ee
where
\be
A=g+3a_{0}^{2}+\sum _{m \geq 1}{6 \over m}a_{m}^{\dagger}a_{m}.
\ee
Sandwiching Eq.(\ref{constraint}) with the Fock vacuum $|0\rangle$ and using
$a_{n}|0 \rangle = \langle 0|a_{n}^{\dagger}=0$, we get
\be
\sigma (\sigma ^{2}+g-{6\zeta (2)\hbar ^{2} \over g+3\sigma^{2}})=0,
\label{sigma}
\ee
where  
$\zeta (2)={\pi ^{2} \over 6}$ for the Riemann zeta function 
$\zeta (k) \equiv \sum _{n \ge 1}{1 \over n^{k}}$.  
The second term in the expression of ${\tilde \alpha}_{1}(n)$  
is derived from the last term of 
the constraint equation (\ref{constraint}).
Both the terms do not contribute to the vacuum expectation value of
$a_{0}$. So we neglect the terms in the following. 
Equation (\ref{sigma}) has non-zero solutions
\be
\sigma^{2}={-2g+\sqrt{g^{2}+3\pi ^{2}\hbar ^{2} } \over 3} ,
\ee 
and one finds a critical coupling $ \lambda _{cr}=24\mu ^{2}/\hbar$ 
from $\sigma=0$. 
In the limit $g \rightarrow -\infty$, we have
$\sigma =\pm(\sqrt{ -g}+{3 \over 2}\zeta (2){\hbar ^{2} \over \sqrt 
{-g^{3}}}+
\cdots ).$
This asymptotic behavior agrees with Ref. \cite{Pinsky2} 
in which certain Feynman
diagrams are calculated to the one-loop level within the framework of 
discretized light-front quantization. One also finds that 
$ \sigma \sim (\lambda - \lambda _{cr})^{\beta}$ with $\beta
=1/2$ near the critical point. The same critical exponent has already been
obtained in Ref.\cite{Pinsky2}, 
but the value $\beta =1/2$ disagrees with the expected value 
$\beta=1/8$ indicated from the Ising model \cite{Domb}.

At $O(\hbar^{2})$, similarly, one finds from the constraint equation 
(\ref{constraint})
\be
{\tilde \alpha} _{2}(n)=-({6 \over nA})^{2}a_{0}a_{n}+{108 \over n^{2}A^{3}} 
a_{0}^{3}a_{n},
\ee
and $\sigma$ satisfies 
\be
\sigma ^{2}+g-{\pi ^{2}\hbar ^{2} \over g+3\sigma^{2}}
+\zeta (3)\hbar ({6\hbar \over g+3\sigma^{2}})^{2}{g \over g+3\sigma^{2}}=0.
\label{sigma2}
\ee
At $O(\hbar ^{2})$  one again finds 
$ \sigma \sim (\lambda - \lambda _{cr})^{1/2}$ near the critical point.

It is not so easy to proceed to the higher order of $\hbar$, because of 
appearance of terms including $a_{m}^{\dagger}a_{m}$ 
and $a_{0}^3$ in ${\tilde  \alpha}_{p}$. However, it is possible to 
set safely $\sigma ^{3} =0$ at the critical point.
So we concentrate on calculating a critical coupling $\lambda_{cr}$.
We then decompose the ${\tilde  \alpha} _{p}(n)$ into 
\be 
{\tilde  \alpha} _{p}(n)=
{ \alpha _{p}(n) \over A^{p}}a_{0}a_{n} 
+ O(a_{m}^{\dagger}a_{m}) + O(a_{0}^{3}) 
\quad (n \geq 1), 
\label{alpha}
\ee
with mere number $ \alpha _{p}(n)$, and neglect the $O(a_{0}^{3})$
terms in the following. 
The second term $O(a_{m}^{\dagger}a_{m})$
contains terms like $ \sum_{m}a_{m}^{\dagger}a_{m}a_{0}a_{n}/A^{p+1}$ and
$\sum_{m,l}a_{m}^{\dagger}a_{m}a_{l}^{\dagger}a_{l}a_{0}a_{n}/A^{p+2}$ that 
vanish
when sandwiched by $\langle 0|$ and $a_{n}^{\dagger}|0\rangle$.
If $\alpha_{p}(n)$ are found, Eq.(\ref{constraint}) yields 
\bea
0&=&\sigma ^{3}+g\sigma-\sum _{p,n \ge 1}{ \alpha_{p}(n) \over n}
\langle 0|A^{-p}a_{0}a_{n}a_{n}^{\dagger} |0\rangle \hbar ^{p}+O(\sigma^{3})
\nonumber  \\
&=&\sigma ^{3}+g\sigma -\sum _{p,n \ge 1}{ \alpha_{p}(n) \over n}
g^{-p}\sigma  \hbar ^{p+1} +O(\sigma ^{3}).
\label{sigma3}
\eea

Now, we calculate  $\alpha_{p}(n)$. It is easy to calculate the commutation
relation (\ref{derivation}) except for the following term
\bea
&\sum _{m \geq 1}&{1 \over m}(a_{m}[a_{0},a_{n}]a_{m}^{\dagger}
+a_{m}^{\dagger}[a_{0},a_{n}]a_{m}) \nonumber \\
= &\sum _{p \geq 1}&{1 \over n}{\alpha _{p}(n) \over A^{p}} a_{0}a_{n} 
\hbar^{p+1}+\sum _{m,p \geq 1}({2 \over m}a_{m}^{\dagger}a_{m}+{\hbar \over 
m} )
{\alpha _{p}(n) \over A^{p}} a_{0}a_{n} \hbar ^{p} \nonumber \\  
 -&\sum _{m,p \geq 1}&{1 \over m}
[[{\alpha _{p}(n) \over A^{p}} a_{0}a_{n},a_{m}],a_{m}^{\dagger}]\hbar ^{p}
+O(a_{m}^{\dagger}a_{m})+O(a_{0}^{3}).
\eea
Collecting $\hbar ^{p}$ terms in Eq.(\ref{derivation}), one obtains
\be
\alpha _{p}(n)=-{6 \over n}\alpha _{p-1}(n)
+ [
{ A^{p-1}}\sum _{m \geq 1}\sum _{k \geq 1}^{p-2}({1 \over m}
[[{\alpha _{k}(n) \over A^{k}} a_{0}a_{n},a_{m}],a_{m}^{\dagger}]\hbar
^{k})
]_{p}    ,  \hspace{5mm} 1 \le p \le 4     
\label{alpha2}
\ee
where $\alpha _{0}(n)=-1$ and the bracket $[ B ]_{p}$ means extracting 
a coefficient $C$ of $a_{0}a_{n}\hbar^{p}$ from the operator 
$B = C a_{0}a_{n}\hbar^{p}+\cdots$.
This recursion relation is true for $1 \le p \le 4 $, 
since it is derived with 
${\tilde \alpha}_{1}$ and ${\tilde \alpha}_{2}$ 
which have no $O(a_{m}^{\dagger}a_{m})$ term.
Using 
\be
 {1 \over A}a_{m}= a_{m}{1 \over A-{6 \over m}\hbar }+O(a_{0}^{2}),
\hspace{5mm} 
 {1 \over A}a_{m}^{\dagger}= a_{m}^{\dagger}{1 \over A+{6 \over m}\hbar 
}+O(a_{0}^{2}),
\ee
one obtains from Eq.(\ref{alpha2})
\bea 
\alpha _{3}(n)&=&{4 \over 3}({6 \over n})^{3}
+{1 \over 3}{6 \over n}\sum_{m \ge 1} ({6 \over m})^{2}, \\ \nonumber 
\alpha _{4}(n)&=&-{7 \over 3}({6 \over n})^{4}
-{5 \over 6}({6 \over n})^{2}\sum_{m \ge 1} ({6 \over m})^{2}
-{1 \over 2}{6 \over n}\sum_{m \ge 1} ({6 \over m})^{3}.
\eea
For $p \ge 5$, on the other hand, the $O (a_{m}^{\dagger}a_{m})$ terms in 
${\tilde \alpha}_{i} \hspace{3mm} (i \ge 3)$ give  
some modifications to the recursion relation. 
In principle, the recursion relation is obtainable also for $p \ge 5$,
but in practice the calculation is very complicated. So we consider
the case of $p \le 4$.


Setting $\sigma ^{3}=0$ in Eq.(\ref{sigma3}), 
we obtain an equation for critical coupling $g_{cr}$, which is valid up to 
$O(\hbar ^{6})$, 
\bea
0&=&6-\zeta (2) ({x \over \pi})^{2}+\zeta (3)( {x \over \pi})^{3}
-({4 \over 3}\zeta (4)+{1 \over 3}\zeta (2)^{2})( {x \over \pi})^{4}
+({7 \over 3}\zeta (5)+{4 \over 3}\zeta (3)\zeta (2))( {x \over \pi})^{5} 
\nonumber \\  
&=&36-x^{2} (1-0.23261x+0.14444x^{2}-0.09913x^{3}),
\eea
where  $x=(6\pi \hbar)/g_{cr}$.
We adopt the Pade approximations \cite{Killing} to an alternating 
series $1-c_{1}x+c_{2}x^{2}-c_{3}x^{3}+\cdots$. This series is
approximated into $[M/N]$, 
which denotes a rational equation of polynomials of degrees $M$ and $N$:
\bea
&[i/1]& \qquad \qquad 1-c_{1}x+\cdots +(-1)^{i}c_{i}{x^{i} \over 1+{c_{i+1} 
\over c_{i}}x} , \\ \nonumber
&[1/2]&  \qquad \qquad  {1+({c_{3}-c_{1}c_{2} \over c_{2}-c_{1}^{2}}-c_{1})x
\over 1+{c_{3}-c_{1}c_{2} \over c_{2}-c_{1}^{2}}x+({c_{3}-c_{1}c_{2} \over 
c_{2}-c_{1}^{2}}c_{1}-c_{2})x^{2}}.
\eea
Lower (upper) bounds of $\lambda _{cr} =(4\mu^{2}x)/\hbar$ 
are obtained from cases of $M+N$=even (odd), 
\bea
&[0/0]& \quad \lambda_{cr}=24 \mu ^{2}/\hbar , \qquad
[1/1] \quad  \lambda_{cr}=28.8 \mu^{2}/\hbar ,  \\ \nonumber
&[0/1]& \quad  \lambda_{cr}=46.0 \mu ^{2}/\hbar, \qquad 
[2/1] \quad  \lambda_{cr}=32.0 \mu ^{2}/\hbar ,\qquad \\ \nonumber
&[1/2]& \quad  \lambda_{cr}=31.1 \mu ^{2}/\hbar.
\eea
Our result is 
$28.8\mu ^{2}/\hbar \le \lambda_{cr} \le 31.1\mu ^{2}/\hbar$,
which is consistent with the equal-time result \cite{PT},
$22\mu ^{2}/\hbar \le \lambda_{cr} \le 55.5\mu ^{2}/\hbar$.

So far the symmetric ordering has been assumed 
in the constraint equation. We now consider 
another possible operator ordering  that does not break the hermitian
nature. Under the ordering the constraint equation becomes 
\bea
0&=&a_{0}^{3}+ga_{0}+\sum _{n \geq 1}{1 \over n}((a_{0}a_{n}a_{n}^{\dagger}
+a_{0}a_{n}^{\dagger}a_{n}+a_{n}a_{n}^{\dagger}a_{0}+
a_{n}^{\dagger}a_{n}a_{0})(1+s)
+(a_{n}a_{0}a_{n}^{\dagger} \nonumber \\&+&
a_{n}^{\dagger}a_{0}a_{n})(1-2s)-3\hbar a_{0}) +\sum _{k,l,m \ne 0}
{\delta_{k+l+m,0} \over \sqrt{ |klm|} }a_{k}a_{l}a_{m}\equiv
 \Phi _{s} (a,a^{\dagger})
\label{constraint2}
\eea
for a real parameter $s$. When $s=0$, this ordering agrees with the
symmetric one. 
Equations corresponding to  
(\ref{sigma3}) and (\ref{alpha2}) are 
\be
0=\sigma ^{3}+g\sigma -\sum _{p,n \ge 1}{ \alpha_{p}(n) \over n}
g^{-p}\sigma  \hbar ^{p+1}(1-2s) +O(\sigma ^{3}),
\ee
\be
\alpha _{p}(n)=-{6 \over n}\alpha _{p-1}(n)
+(1-2s)[{ A^{p-1}}\sum _{m \geq 1}\sum _{k \geq 1}^{p-2}({1 \over m}
[[{\alpha _{k}(n) \over A^{k}} a_{0}a_{n},a_{m}],a_{m}^{\dagger}]\hbar 
^{k})]_{p}    ,  
\label{alpha3}
\ee
for $ 1 \le p \le 4 $.  Since the double bracket term 
in Eq. (\ref{alpha3}) is $O(\hbar^3)$, 
$\alpha _{1}$ and $\alpha _{2}$ do not depend on $s$. 
At $O(\hbar)$ we have 
\be
\sigma (\sigma ^{2}+g-{\pi ^{2}\hbar ^{2} \over g+3\sigma^{2}}(1-2s))=0,
\ee
so that $\lambda_{cr}=24\mu ^{2}/(\hbar \sqrt{1-2s})$ and $\beta=1/2$ 
for $ s < 1/2 $. The spontaneous breaking does not occur for $ s \ge 1/2$.
When $ -0.095 \le s \le 0.407 $, the critical coupling calculated here 
reproduces the equal-time one 
$22\mu ^{2}/\hbar \le \lambda_{cr} \le 55.5\mu ^{2}/\hbar$.     
If the equal-time calculation yields 
$\lambda_{cr}$ more accurately, one can find which $s$ reproduces 
the $\lambda_{cr}$ and can know whether the symmetric ordering is the only
ordering which reproduces the equal-time result.
While $\lambda_{cr}$ depends on $s$, $\beta$ does not. This
property persists for the higher order of $\hbar$.
The present value of $\beta$ is still different from the expected value 
$1/8$.

In this paper, we propose a systematic way of obtaining 
$\lambda_{cr}$ and $\beta$ within the framework of 
discretized light-front quantization. 
When the symmetric ordering is taken in the constraint equation, 
this formulation yields a reasonable value
of $\lambda_{cr}$. The calculated $\lambda_{cr}$ has
an error much smaller than that of the equal-time calculation. 
This light-front formulation, however, does not reproduce the 
expected critical behavior $\sigma \sim (\lambda - \lambda _{cr})^{1/8}$.
As long as the expansion (\ref{expansion}) of $[a_{0},a_{n}]$ 
is truncated at a finite order of $\hbar$, 
the non-zero solutions of $\sigma$ obey an equation in which the higher
order terms are added to Eq. (\ref{sigma2}). 
It is unlikely that in the equation the first term $\sigma^2$ is
canceled by the higher order terms. So the unsatisfactory critical
behavior $\sigma \sim (\lambda - \lambda _{cr})^{1/2}$ 
does not seem to break. This behavior is not changed by taking 
another operator ordering (\ref{constraint2}). This is an important
problem to be solved. The first thing which we must do 
is to find the commutation relation $[a_{0},a_{n}]$ without the
truncation.


\begin{thebibliography}{99}
\bibitem{Chang}
      S.J. Chang, R.G. Root and T.M. Yan, Phys. Rev. {\bf D7},
      1133(1973);  
      S.J. Chang and T.M. Yan, Phys. Rev. {\bf D7}, 1147(1973);
      T.M. Yan, Phys. Rev. {\bf D7}, 1760(1973);
      T.M. Yan, Phys. Rev. {\bf D7}, 1780(1973).
\bibitem{Brodsky}
      H.-C.Pauli and S.J.Brodsky, Phys.Rev. {\bf D32}, 1993(1985);
      H.-C.Pauli and S.J.Brodsky, Phys.Rev. {\bf D32}, 2001(1985);
      T. Eller, H.C. Pauli and S. Brodsky, Phys. Rev. {\bf D35}, 1493(1987);
      T. Eller and H.C. Pauli, Z. Phys. {\bf C42}, 59(1989); 
      K. Hornbostel,S. Brodsky and H.Pauli, Phys. Rev. {\bf D41},
      3814(1990). The heading is mislabeled 
         in Table I: $M/g$ should be replaced 
         by $M^2/(g^2/\pi + m^2)$. This is informed by 
         Dr. Hornbostel. 
\bibitem{Burkardt}
     M. Burkardt, Nucl. Phys. {\bf A504}, 762(1989). 
\bibitem{ours}
         K. Harada, T. Sugihara, M. Taniguchi and M. Yahiro,
         Phys. Rev. {\bf D49}, 4226(1994);
         T. Sugihara, M. Matsuzaki and M. Yahiro,
         Phys. Rev. {\bf D50}, 5274(1994).
\bibitem{MY}
         T. Maskawa and K. Yamawaki, 
         Prog. Theor. Phys. {\bf 56}, 270(1976). 
\bibitem{Yamawaki}
         N. Nakanishi and K. Yamawaki,
         Nucl. Phys. {\bf B122}, 15(1977); 
         Y. Kim, S. Tsujimaru and K. Yamawaki,
         Phys. Rev. Lett. {\bf 74}, 4771(1995).
\bibitem{Hornbostel}
         K. Hornbostel, Phys. Rev. {\bf D45}, 3781(1992). 
\bibitem{PT}
         S.J. Chang, Phys.Rev. {\bf D13}, 2778(1976);
         J. Abad, J. G. Esteve and A. F. Pacheco, Phys.Rev. {\bf D32}, 
         2729(1985);
         M. Funke, U. Kaulfuss and H. Kummel, Phys.Rev. {\bf D35}, 
         621(1987);
         H.Kroger, R. Girard, and G. Dufour,Phys.Rev. {\bf D35},
         3944(1987).
\bibitem{Heinzl}
         T. Heinzl, S. Krusche, S. Simburger and E. Werner, 
         Z. Phys.{\bf C56}, 415(1992); Phys.Lett.{\bf B272}, 54(1991); 
         Phys.Lett. {\bf B275}, 410(1992);
         T. Heinzl, C. Stern, E. Werner and B. Zellermann, 
         Z. Phys. {\bf C72}, 353(1996). 
\bibitem{Harin}
         A. Harindranath and J. P. Vary, Phys. Rev. {\bf D36},
         1141(1987)1141; Phys. Rev.{\bf D37}, 1076(1988);
         D. G. Robertson, Phys. Rev. {\bf D47}, 2549(1993);
         M. Burkardt, Phys. Rev. {\bf D47}, 4628(1993).
\bibitem{Pinsky1}
         C. M. Bender, S. S. Pinsky and B. van de Sande, Phys. Rev.
         {\bf D48}, 816(1993);
         S. S. Pinsky and B. van de Sande, Phys. Rev. {\bf D49},
         2001(1994).
\bibitem{Pinsky2}
         S.Pinsky, B. van de Sande and J. R. Hiller, Phys. Rev. 
         {\bf D51}, 726(1995).
\bibitem{Xu}
         X. Xu and H. J. Weber, Phys. Rev. {\bf D52}, 4633(1995).
\bibitem{Pinsky3}
         C. M. Bender, L. R. Mead and S. S. Pinsky, 
         Phys. Rev. Lett. {\bf 56}, 2445(1986).
\bibitem{Domb}
         C. Domb and M. S. Green, {\it Phase Transitions and Critical
         Phenomena, Vol.6} (Academic, London, 1976).
\bibitem{Killing}
         J. Killingbeck, {\it Microcomputer Quantum Mechanics}
         (Adam Hilger, Bristol, 1983). 
\end{thebibliography}
\end{document}